\documentstyle[psfig]{caosp}

\begin{document}
\pubyear{1998}
\volume{27}
\firstpage{184}
\htitle{CP stars: Photometric calibrations of luminosity using Hipparcos
data}
\hauthor{F. Figueras {\it et al.}}
\title{CP stars: Photometric calibrations of luminosity using Hipparcos data}
\author{F. Figueras \inst{1} \and X. Luri
\inst{1,} \inst{2}  \and A.E. G\'omez \inst{2} \and
J. Torra \inst{1} \and C. Jordi \inst{1} \and M.O. Mennessier \inst{3}
\and A. Domingo \inst{1} \and E. Masana \inst{1} \and
S. Grenier \inst{2} \and F. Blasi \inst{1}} 

\institute{Departament d'Astronomia i Meteorologia. Univ. de Barcelona, Spain
\and Observatoire de Paris-Meudon, DASGAL, France
\and GRAAL. Universit\'e de Montpellier II, France}
\maketitle
\begin{abstract}
The application of the Str\"omgren photometric 
luminosity calibrations  to  different types of CP stars is 
reexamined in the light of the new Hipparcos data. 
A first attempt is made to use the LM statistical 
parallax method (Luri {\it et al.}, 1996) --  based on the 
maximum likelihood principle -- to  obtain a calibration  
of the absolute magnitude  as a function of two Str\"omgren 
colour indices,  thus reflecting effective temperature and 
evolution.  Its application to a sample of Si stars
and to a sample of normal main sequence stars in the 
same spectral range allows us to compare the calibrations
obtained and to discuss the position of Si stars 
in the HR diagram.  Additionally, a sample 
of {\it bonafide}, spectroscopically  selected Am stars together with normal 
main sequence stars are used to derive a new absolute magnitude 
calibration for late A-type main sequence stars, taking into 
account the effects of evolution, metallicity and stellar rotation. 

\end{abstract}
\section{Introduction}
To deal with luminosity calibrations and to obtain a good exploitation 
of the high precision parallax data obtained by 
Hipparcos, it is required to use both, robust statistical methods, 
capable to take into account 
physical characteristics of stars (i.e. evolution, metallicity, rotational
effects, etc.),
and a well defined spectroscopic sample. 
Taking advantage of the fact that Str\"omgren photometry
has proved to be a powerful tool for characterizing the physics of
main sequence early type stars, some preliminary work is presented here for
application to Si and Am stars.

\begin{figure}[th]
\centerline{
\psfig{figure=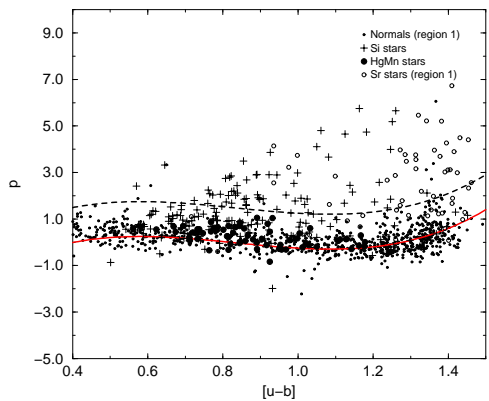,height=6.2cm}}
\caption{Detection of hot CP stars using the $\Delta p$ parameter:
$\Delta p = p -p_o = 1.5$ (dashed line), $p_o = f([u-b])$ being the
relation for normal stars (full line)}
\label{fig1}
\end{figure}

Our sample contains all CP stars included in the Hipparcos Catalogue 
(ESA, 1997) having good spectroscopic information on peculiarity and 
complete Str\"om\-gren photometry in the 
Hauck and Mermilliod (1996) compilation. The
selection of the stars according 
to their spectral types
is fully explained in G\'omez et al. (1998, this colloquium). For comparison,
a sample of normal main sequence stars in the range B0-A9 V, selected from 
the Hipparcos Survey  (G\'omez et al., 1997), 
has been  also used. Table~\ref{t1} 
shows the stars in each CP group (the sample is reduced 
to less than 50 \% when Str\"omgren photometry is required).
\begin{table}[t]
\small
\begin{center}
\caption{CP stars with $uvbyH_{\beta}$ photometry.}
\label{t1}
\begin{tabular}{|c|c|c|c|} \hline
\hline
& Selection from & with $uvbyH_{\beta}$ & Range of \\
& G\'omez et al. (1997a) & Hauck \& Mermilliod & SP \\
&  & (1996)   &   \\
\hline
\hline
Si             & 440 & 173 & B5-A3 \\
HgMn           &  76 & 69 & B7-A0 \\
Sr-Cr-Eu       & 378 & 172 &B8-A9  \\
Am             & 1059 & 533& A0-A9 \\
Normal B0-A9 V & 3460 & 1589 & B0-A9 \\
\hline
\end{tabular}
\end{center}
\end{table}
Using the different samples, a 
test of the capability of the $uvbyH_{\beta}$ 
system to detect CP stars has been performed.  The $([m_1], [c_1])$ plane
has been classically accepted as  
the most discriminant plane in the Str\"omgren system to separate CP
from normal stars. Thus for example, Crawford (1979) established the criterion 
$\delta m_1 \lid -0.020$ to separate Am stars from A3-A9 normal main 
sequence stars. In contrast to Abt (1984), who concluded that about
75 \% of the Am stars could be photometrically classified as such, we
 obtain that only 55 \% of the classical Am 
(Sp(k)-Sp(m) $\gid$ 5) and 33 \% of the proto-Am  (Sp(k) - Sp(m) $<$ 5)
can be detected using only 
Str\"omgren colours. Philip et al (1976) criteria $(E(b-y) \lid$ -0.040)
to detect hot CP stars is found to be unable to acomplish
the objective (only 2 \% of Si stars and 5 \% of Sr stars has been detected).
Taking into account that peculiar spectral features alter all Str\"omgren 
indices,
Masana et al. (1998) established a new criterion of detection
defining a $\Delta p$ parameter 
-- a linear combination of several $uvbyH_{\beta}$ indices obtained through
Multiple Discriminant Analysis. Applying this criterion to the hot stars in
the present sample (see Figure~\ref{fig1}) we photometrically classify as 
peculiar 31 \% of the Si stars and 
56 \% of hot Sr-Cr-Eu stars, 
whereas no HgMn star is detected. Furthermore,
as indicated by Masana et al. (1998), the reddening decreases $\Delta p$, so
reddened hot CP stars may be seen as non-reddened normal stars, thus decreasing
its capability for detection. We can conclude that  the 
Str\"omgren system is suitable but less powerful for a photometric 
detection of peculiarities
than the $\Delta a$ system (Maitzen, 1976; specifically designed to
measure characteristic features on the spectra of peculiar stars) 
or even the  $\Delta (V_1 - G)$ combination 
of the Geneva indices  (Masana et al., 1998).

A crucial point to attack the problem of the  $M_v$ calibration of CP stars 
is the reddening correction. For CP2 stars, the peculiarities in the 
spectra result in bluer colour indices, the
$(b-y)$ and $c_1$ indices being smaller, thus leading to
underestimate their reddening when treating them as normal stars.
In agreement with Adelman (1980) and Maitzen (1980), Masana et al. (1998) 
derived a correction  to the $E(b-y)$ obtained when using standard relations 
valid for normal stars as a function of the $\Delta p$ parameter 
($\Delta E(b-y) = -0.001 + 0.008 \Delta p$). Even applying this correction
to the stars with $\Delta p \geq 1.5$,  
we obtain that 57 \% of Si 
stars and 38 \% of Sr-Cr-Eu (early region) stars have $E(b-y) < 0$ , compared
with only 9 \% of normal main sequence stars in the same spectral
range with negative excess (using Crawford's (1978) standard relations).
The problem is maintained in the intermediate region (Grosb{\o}l's (1978) 
standard relation): 56\% of Sr-Cr-Eu stars have negative excess compared to
37 \% of normal main sequence stars. In the late region (A3-A9), where
Crawford's (1979) standard relation is considered, 
the overestimation of reddening is also clear for Sr-Cr-Eu 
(18 \% with $E(b-y) <0$  compared with
37\%  for normal stars). The same excess distribution is obtained
for Am and normal A3-A9 stars (35 \% and 37 \% of stars with $E(b-y)<0$ 
respectively). Having in mind these biased photometric reddening 
determinations when peculiarities are present, we preferred to use the
interstellar absorption model by Arenou et al. (1992) in the LM method, 
which gives reddening corrections 
as a function of the star position $(r,l,b)$.
Improvement of the interstellar absorption models using the new Hipparcos data 
is needed.
\section{Testing the current available Mv photometric calibrations} 
The photometric distances computed following the algorithms of Crawford (1978), 
Balona \& Shobbrook (1984),
Jakobsen (1985) and Eggen (1974) for the stars belonging to 
the early region, are compared in Table~\ref{t2}
with the distances derived directly from Hipparcos parallaxes for
Si, Hg-Mn and Sr-Cr-Eu stars as well as for B4-B9 V normal stars.
Only stars
with good parallax, $\Delta \pi / \pi \leq 30 \%$, have been  used. All these
\begin{table}[h]
\small
\begin{center}
\caption{Mean differences between Hipparcos and Str\"omgren photometric
distances (only stars with $\Delta \pi / \pi \leq 30 \%$ are considered),
Units: parsecs.}
\label{t2}
\begin{tabular}{|l|c|c|c|c|} \hline
\multicolumn{5}{|c|} {} \\
\multicolumn{5}{|c|} {\bf Early Region} \\
\hline
{\bf Calibration}& {\bf Si} & {\bf Hg-Mn} & {\bf Sr-Cr-Eu} & {\bf B4-B9 V}\\
                 &          &             &     hot        & normal \\
                 &  (107 *) & (67 *)      &   (47 *)       & (361 *)\\
\hline
Crawford (1978)       & -16 $\pm$ 67 & +3 $\pm$ 48 & +6 $\pm$ 40& +1 $\pm$ 60\\ 
Balona \& Shobbrook (1984)& -28$\pm$ 84&+0 $\pm$ 50& +3 $\pm$ 48& -1 $\pm$ 60\\ 
Jakobsen (1985)        & -15 $\pm$ 65 & -4 $\pm$ 51&-14 $\pm$ 55& -5 $\pm$ 61\\
Eggen (1974)             & -61 $\pm$ 97 &             &        &-19 $\pm$ 69\\ 
\hline
\multicolumn{5}{|c|} {} \\
\multicolumn{5}{|c|} {\bf Intermediate and late Regions} \\
\hline
{\bf Calibration}&{\bf Sr-Cr-Eu }&{\bf Sr-Cr-Eu}&{\bf A0-A2 V} & {\bf A3-A9 V}\\
                 &  intermediate   &  late          &  normal     & normal \\
                 &  (26 *)         & (36 *)         &   (338 *)   & (376 *)\\
\hline
Str\"omgren (1966) & +23 $\pm$ 43  &                & +6 $\pm$ 31 &         \\
Guthrie (1987) &      & +62 $\pm$ 84   &             & +10 $\pm$ 38 \\
\hline
\end{tabular}
\end{center}
\end{table}
algorithms, based on normal main sequence stars, rely on the fact that
the $\beta$ index is primarily a function of luminosity and that
$c_o$ and  $[u-b]$ are related to temperature in the early region. As the 
$c_o$ index of CP2 stars falls below those of 
normal stars for a given spectral type (Preston, 1974), systematic trends
in the $M_v$ derivation 
can be present when using the calibrations by Crawford (1978),
Balona \& Shobbrook (1984) and Jakobsen (1985). 
 Napiwotzki et al. (1993) and M\'egessier (1988) pointed out that the
$[u-b]$ index is slightly more sensitive to temperature than $c_o$ and can
be accepted as a good indicator of temperature for Si stars. 
Eggen (1974) used this index but his calibration shows strong 
systematic trends. 
 On the other hand,
the $\beta$ index is also affected 
by the peculiarities in the spectra, as pointed out by Hauck (1975).

From Table~\ref{t2} we see that, due to the above-mentioned effects and to the 
underestimation of reddening (when computed using Str\"omgren 
photometry), 
the photometric distances of Si stars
have always been  overestimated. No significant differences are found
neither for Hg-Mn stars (as pointed out by Str\"omgren (1966) these stars
behave as normal stars as far as the $c_1$, $m_1$ and $\beta$ indices
are concerned) nor for B4-B9 V normal stars. The standard
deviation obtained ($\sim $ 60 pc for normal stars) agrees with a relative
error in photometric distances of about 30\%. 

Although the  intermediate and late regions contain few Sr-Cr-Eu stars, 
Table~\ref{t2} shows  
that  the photometric distances 
are underestimated when classical calibrations for normal stars
are used. Again, the overestimation of reddening for these stars can
contribute to this effect.
\section{A new calibration for Si stars using the LM method}
The LM (Luri et al., 1996), based on the
Maximum Likelihood principle, allows a simultaneous determination of
a luminosity calibration, kinematical characteristics and spatial
distribution of a given sample. In a different approach that the one
presented by G\'omez el al. (1997a) in this colloquium, we undertake
a derivation of a calibration of absolute magnitude of Si stars
and B4-B9 V normal main sequence stars by  
parametrizing $M_v$ as a linear
combination of Str\"omgren colour indices ($Q$), the observational entries
being then ($\alpha, \delta, m, \pi, \mu, V_r, Q$). No selection function
for colours is imposed ($S(Q) = 1$), and an asymmetrical gaussian 
distribution of $Q$ around a maximum ($C_o$) colour index is modelled.
 For a fixed value of $Q$, a gaussian distribution for $M_v$ is assumed:
\begin{equation}
\Phi_M (M | Q) = {\large e}^{ - \frac{1}{2} (\frac{M-M_0(Q)}{\sigma_M})^2}
\end{equation}
and as a first attempt we consider the dependence of $M_v$ on the 
effective temperature:
\begin{equation}\label{r2}
M_0(Q) =  p_1 + p_2 ( log_{10}([u-b] + 0.2))
\end{equation}
$p_1$ and $p_2$ are simultaneously determined together with 
the absolute magnitude dispersion ($\sigma_{M_v}$) ,  the parameters
defining the velocity ellipsoid (mean $(U,V,W)$ and dispersions
($\sigma_U, \sigma_V, \sigma_W$)) and
the scale height ($Z_o$).
We use the $[u-b]$ index as explained above
and a logarithmic function allowing us to solve the non-linearity
of the main-sequence in this spectral range (Balona and Shobbrook,
 1984). The method, able to identify and characterize distinct physical
groups in inhomogeneous samples, has been applied considering the presence
of two groups. Results for Si stars
indicate the existence of 
a main group (containing 98 \% of the sample) and a secondary group
(2\%) probably composed of misclassified objects and/or 
high velocity stars. Approximately
the same percentage of classification is obtained when normal stars are
considered. The relations obtained for the main groups are:
$M_0 =  -0.10 + 4.02 ( log_{10}([u-b] + 0.2))$
for normal B4-B9 V (416 stars) and 
$M_0 =  0.01 + 3.58 ( log_{10}([u-b] + 0.2))$
for Si stars (149 stars).

One of the major advantages of the Str\"omgren system, compared with other 
photometric systems for early type stars, is its capacity to account for 
evolutionary effects through the $\beta$ index. A first attempt was made
to incorporate directly in equation~(\ref{r2}) a term depending on $\beta$
but the high correlation between $\beta$ and $[u-b]$ did not allow
to obtain stable values for the  $p_i$ coefficients. As a preliminary attempt
we have applied the LM method in two steps: once the LM algorithm has been
applied using equation~(\ref{r2}), we seek for a  second solution considering:
\begin{equation}\label{r5}
M_0 - M_0 (1^{srt} \hspace{0.1cm}step)=  p_1'+ p_2'(\beta)
\end{equation}
For the main group (98 \% of stars) the following relations have been obtained:
\begin{equation}
M_0 =  -16.44 + 4.02 ( log_{10}([u-b] + 0.2)) + 5.83 \beta
\end{equation}
for normal B4-B9 V  (416 stars), with
     $\sigma_M = 0.60 $ mag (in the  1$^{rst}$ step : 0.70 mag),
     $(U_0. V_0, W_0)$ = (-12,-15, -7) km/s,
     $(\sigma_U, \sigma_V, \sigma_W)$ = (9,11,5) km/s,
     $Z_0$ =  63 pc, and
\begin{equation}
M_0 =  -21.70 + 3.58 ( log_{10}([u-b] + 0.2)) + 7.86 \beta
\end{equation}
for Si stars (149 stars), with
     $\sigma_M = 0.77 $ mag (in the  1$^{srt}$ step : 0.83 mag),
     $(U_0. V_0, W_0)$ = (-15,-15, -8) km/s,
     $(\sigma_U, \sigma_V, \sigma_W)$ = (9,10,6) km/s,
     $Z_0$ =  66 pc.
\begin{figure}[hbt]
\centerline{
\psfig{figure=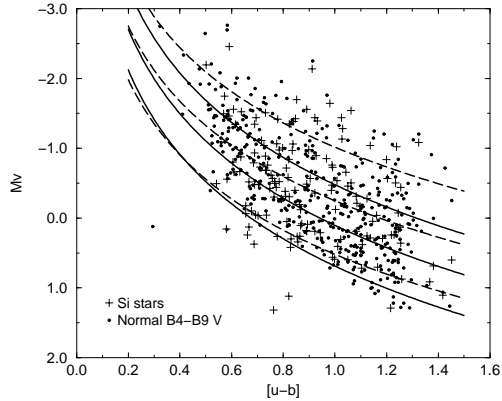,height=6.2cm}}
\caption{Application of the LM method in two steps in which the evolutionary
effects are acounted for. Full lines: relations obtained for normal 
B4-B9 V stars plotted at $\beta$ = 2.6, 2.7 and 2.8 (from top  
to bottom of the figure). Dashed lines: the same for Si stars}
\label{fig3}
\end{figure}
Relations (4) and (5), plotted in figure~\ref{fig3} for different 
$\beta$ values, are defined for the whole
population, so
the mean relations are corrected for the Malmquist bias which is present in the 
samples used, which are limited by apparent magnitude.
We can see a different behaviour of the $\beta$ index
as indicator of luminosity for normal
and Si stars, the differences being more pronounced when decreasing 
the effective temperatures. 
Nonetheless, from the position of individual stars in the HR diagram 
computed with this
solution, 
we can conclude that the evolutionary state of Si stars is  nearly 
the same as that of
normal main sequence stars.
More work will be devoted in the near future to finding,
through Principal Component Analysis,
a suitable proper combination
of colour indices that could allow a simultaneous determination of all the
$p_i$ parameters using the LM method.
\section{A new calibration for Am stars using the BCES method}
A different  statistical treatment is used for  the
determination of the luminosity calibration in the late  Str\"omgren's
region, taking into account the dependence of  $M_v$
on  effective  temperature,  evolution,  metallicity and rotation.
BCES fitting algorithm (Bivariate Correlated Errors and  intrinsic
Scatter, Akritas \& Bershady, 1996)  is  an  extension of the
ordinary least squares estimator for two variables. It allows
the consideration of measurement errors on both variables and  the
use of  correlated errors between them, and also takes into account unknown
intrinsic  scatter  and  the  possibility  that  the  size  of the
measurement error depends on the variables.

This algorithm has been applied to a carefully selected sample  of
97  Am  stars (North  et  al.,  1997)  and  very accurate 
Hipparcos trigonometric
parallaxes  ($\Delta  \pi  /  \pi  \lid  0.14$).  The 
photometric data has been taken from Hauck and Mermilliod (1996) and the
projected  rotational  velocity  $V  sin  i$  from Abt and Morrell
(1995).  Individual correction for binarity has been applied  whenever
possible (30 \% of the  sample) and a mean correction of  0.2 mag
for SB1 binaries and  stars with variable radial velocity has been considered, 
leaving 29 stars without
correction.  No correction for interstellar absorption has been applied,
after  realizing  that it is meaningless when comparing it with the
uncertainties in  the binarity  correction, and considering that the
$E(b-y)$ computed  using  Crawford's (1979) standard  relation is
distributed around a mean value of 0.0014 mag.

To  estimate  the  contribution  of  the  stellar  rotation  and the
blanketing effects (which are highly correlated in this spectral range)
on the photometric $M_v$ calibration, it is necessary
to have a large range of $V\sin i$ and $\delta m_1$.
For this reason, the calibrations  have been obtained by joinig  to
the sample of Am stars a sample of normal A3-A9 V stars  carefully
selected to  be non-binaries and  non-peculiar. The quality and
source of its trigonometric  parallaxes, photometry and $V\sin i$
is the same as for the Am stars.

An  evaluation  of   the  existing  photometric calibrations  of
luminosity through their comparison with the Hipparcos data, allows  us
to conclude that  Crawford's (1979) calibration  is systematically
shifted by 4 $mas$ in parallax ($\pi_{Hipp} < \pi_{Craw}$) for Am stars,
but shows no systematic trend for normal A3-A9 stars.  The inclusion
of $\delta  m_1$ and $V\sin i$ terms proposed by  Guthrie
(1987)  reduces Crawford's (1979) shift for Am stars but
introduces a systematic deviation for A3-A9 normal stars.

The BCES algorithm has been applied to the sample of Am plus A3-A9
normal stars in an iterative procedure, taking into account all the
errors in the input parameters. The obtained calibration using
Hipparcos data is (Domingo, 1998):\\
\newpage
\noindent
\[ M_v=(12.16\pm 0.03) -(3.5\pm 0.7) \beta -(8.3\pm 0.3)  \delta c_1 + \]
\[ + (6.1\pm 1.0) \delta m_1 + (7.1\pm 1.1) \cdot 10^{-6} (V  \sin i )^2 \]\\
\noindent The unweighted residuals obtained in $M_v$ are 0.25 mag
for A3-A9 V normal stars and 0.36 mag for Am stars, smaller in all
cases than for the existing photometric calibrations in this
spectral region.

\acknowledgements
This work has been supported by CICYT under contract PB95-0180, and by the
PICS program (CNRS PICS 348, CIRIT).


\end{document}